\documentstyle[12pt,aasms4]{article}

\lefthead{SENGUPTA}
\righthead{MAGNETIC FIELD DECAY IN NEUTRON STARS}
\begin{document}

\title{EVOLUTION OF CRUSTAL MAGNETIC FIELDS IN ISOLATED NEUTRON STARS : 
COMBINED EFFECTS OF COOLING AND CURVATURE OF SPACE-TIME}

\author{SUJAN SENGUPTA}
\affil{Mehta Research Institute of Mathematics and Mathematical Physics \\
 Chhatnag Road, Jhusi, Allahabad 211 019, India. E-mail: sujan@mri.ernet.in}

\begin{abstract}

The ohmic decay of magnetic fields confined within the crust of neutron stars
is considered by incorporating both the effect of neutron star cooling
and the effect of space-time curvature produced by the
intense gravitational field of the star. For this purpose a stationary and static
gravitational field has been considered with the standard as well as
the accelerated cooling models of
neutron stars. It is shown that general relativistic
effect reduces the magnetic field decay rate substantially. At the late 
stage of evolution when the field decay is mainly determined by the
impurity-electron scattering, the effect of space-time curvature suppresses
the role of the impurity content significantly and reduces the decay
rate by more than an order of magnitude. Even with a high impurity content
the decay rate is too low to be of observational interest if the 
accelerated cooling model along with the effect of space-time curvature
is taken into account. It is, therefore, pointed out that if a decrease
in the magnetic field strength by more than two orders of magnitude from its initial value
is detected by observation then the existence of
quark in the core of the neutron star would possibly be ruled out. 

\end{abstract}


\section{INTRODUCTION}

The magnetic field evolution in neutron stars has been the subject of much
discussion over the years both in the observational and in the theoretical
context. Recent observations of pulsars and their statistical analysis seem to
imply that isolated pulsars do not undergo significant
magnetic field decay during their life times (Wakatsuki et
al 1992). The possible physical processes through which
the magnetic field in the crust of isolated neutron stars may decay are
 ambipolar diffusion, Hall drift, Ohmic dissipation, etc.
The reviews of Lamb (1991), Chanmugam (1992) and Phinney \& Kulkarni (1994)
provide a reasonable account of our present understanding on the decay of
magnetic fields in isolated neutron stars.

Sang \& Chanmugam (1987) demonstrated that the Ohmic decay of the dipole
magnetic fields is not exponential. Considering magnetic field configurations
which are initially confined to a small part of the crust and which vanish in
the stellar core, several authors (Chanmugam \& Sang 1989; Urpin \& Muslimov
1992; Urpin \& Van Riper 1993) found that the relatively low electrical
conductivity of the crustal matter causes the decay times too short
to be of observational interest if the impurity concentration is high.
On the other hand, Pethic \& Sahrling (1995) showed that if the field does not
vanish in the core then the shortest possible decay time becomes about two
orders of magnitude larger than the characteristic time-scale for decay
when the field vanishes in the core.
 Bhattacharya \& Datta (1996)
have studied the ohmic diffusion of magnetic flux expelled from the
super-conducting core to the crust owing to spindown-induced field evolution
and claimed that the result obtained by them is in agreement with the ohmic
time-scale that appears to be required to explain the observed magnetic
field strengths of isolated and binary neutron stars in the spindown-induced
magnetic flux expulsion scenario.

 Recently Sengupta (1997) demonstrated
the significant contribution of space-time curvature in reducing the decay rate
of the crustal magnetic fields in isolated neutron star by assuming a
spherically symmetric stationary gravitational field. However, the cooling of
neutron star has not been incorporated in that work for the sake of simplicity.
  
   In this paper the Ohmic dissipation of the crustal magnetic field has been
investigated by incorporating both the effects of neutron star cooling and
the space-time curvature produced by the intense gravitational field of the
star. The paper is organized in the following ways: in Section 2, the relevant
equations are presented. In Section 3, the adopted neutron star model,
the initial configurations of the magnetic
field and the electrical conductivity have been discussed. The results are
outlined in Section 4 and finally in Section 5 specific conclusions are made.

\section{MAGNETIC FIELD EVOLUTION IN CURVED SPACE-TIME}

 Assuming hydrodynamic motions to be negligible and the anisotropy of the electrical
conductivity of the crustal material is small, the induction equation in flat space-time can be
written as:

  \begin{equation}\label{FLAT}
\frac{\partial{\bf B}}{\partial t}=-{\bf \nabla} \times \left(\frac{c^2}{4\pi
\sigma}{\bf \nabla}\times{\bf B}\right),
\end{equation}
where $\sigma$ is the electrical conductivity.

The covariant form of Maxwell's equations are given by:

\begin{equation}\label{max1}
\frac{1}{\sqrt{-g}}\frac{\partial}{\partial x^{\nu}}(\sqrt{-g}F^{\mu\nu})=
-\frac{4\pi}{c}J^{\mu}
\end{equation}
and
\begin{equation}\label{max2}
\frac{\partial F^{\mu\nu}}{\partial x^{\lambda}}+
\frac{\partial F^{\nu\lambda}}{\partial x^{\mu}}+
\frac{\partial F^{\lambda\mu}}{\partial x^{\nu}} = 0.
\end{equation}

The generalized Ohm's law can be written as :
\begin{equation}\label{ohm}
J^{\mu}=\sigma g^{\mu\nu} F_{\nu\lambda}u^{\lambda},
\end{equation}
where $F_{\mu\nu}$ are the components of the electro-magnetic field tensor,
$J^{\mu}$ are the components of the 
four-current density, $u^{\mu}$ are the components of the four velocity of
the fluid, $g_{\mu\nu}$ are the components of space-time metric that
describes the background geometry and $g={\rm det}|g_{\mu\nu}|.$ 
Here and afterwards Latin indices run over spatial co-ordinates only whereas
Greek indices run over both time and space co-ordinates.

If a stationary gravitational field is taken into account, then using equations
 (\ref{max1}), (\ref{max2}) and (\ref{ohm}) (neglecting the displacement
current and taking $u^i=0$) the corresponding 
induction equation in curved space-time can be derived as:
\begin{equation}\label{GTR1}
\frac{\partial F_{kj}}{\partial x^0}=
\frac{\partial}{\partial x^k}\left[\frac{c}{4\pi}
\frac{1}{\sqrt{-g}}g_{\mu j}\frac{1}{\sigma u^0}
\frac{\partial}{\partial x^i}(\sqrt{-g}F^{\mu i})\right]
- \frac{\partial}{\partial x^j}\left[\frac{c}{4\pi}\frac{1}{\sqrt{-g}}
g_{\mu k}\frac{1}{\sigma u^0}\frac{\partial}{\partial x^i}(\sqrt{-g}
F^{\mu i})\right].
\end{equation}

For the description of the background geometry I consider the exterior
Schwarzschild metric which is given by
\begin{equation}\label{sch}
ds^2=(1-\frac{2m}{r})c^2dt^2-(1-\frac{2m}{r})^{-1}dr^2-r^2(d\theta^2+
\sin^2\theta d\phi^2),
\end{equation}
where $m=MG/c^2$, $M$ being the total gravitational mass of the core.
The justification for adopting the exterior Schwarzschild metric is provided
in Sengupta (1997). Since,
the crust consists of less than a few percent of the total gravitational mass,
$M$ can be regarded as the total mass of the star. The non-zero components of
the orthonormal tetrad $\lambda^{\gamma}_{(\alpha)}$ of the local Lorentz frame for
the Schwarzschild geometry is given by
$$\lambda^t_{(t)}=(1-\frac{2m}{r})^{-1/2}, \; \lambda^r_{(r)}=
(1-\frac{2m}{r})^{1/2}, \; \lambda^{\theta}_{(\theta)}=1/r, \;
\lambda^{\phi}_{(\phi)}=1/r\sin\theta.$$

If $F_{(\alpha\beta)}$ are the components of the electro-magnetic field tensor in
a local Lorentz frame, then the components of the electro-magnetic field tensor
$F_{\gamma\delta}$ are defined in the curved space-time through the relation:
\begin{equation}\label{TET}
F_{(\alpha\beta)}=\lambda^{\gamma}_{(\alpha)}\lambda^{\delta}_{(\beta)}
F_{\gamma\delta}.
\end{equation}

Using the metric given in equation (\ref{sch}), equation (\ref{GTR1})
can be reduced to 
\begin{equation}\label{GTR1a}
\frac{\partial F_{kj}}{\partial x^0}= \frac{c}{4\pi}
\left[\frac{\partial}{\partial x^k}\left\{
\frac{1}{r^2\sin\theta}g_{lj}\frac{1}{\sigma u^0}
\frac{\partial}{\partial x^i}(r^2\sin\theta F^{li})\right\}-
 \frac{\partial}{\partial x^j}\left\{\frac{1}{r^2\sin\theta }
g_{lk}\frac{1}{\sigma u^0}\frac{\partial}{\partial x^i}(r^2\sin\theta F^{li})
\right\}\right].
\end{equation}

Following the convention, I consider the decay of
a dipolar magnetic field which has axial symmetry so that the vector potential
${\bf A}$ may be written as $(0,0,A_{\phi})$ in spherical polar co-ordinates
 where $A_{\phi}=A(r,\theta,t).$ Since the hydrodynamic motion is negligible
 so $u^i= dx^i/ds = 0$ and the metric gives
$$u^0=(1-\frac{2m}{r})^{-1/2}.$$

Therefore, from equation (\ref{GTR1a}) we obtain
\begin{equation}
\frac{\partial F_{r\phi}}{\partial t} = \frac{c^2}{4\pi}
 \frac{\partial}{\partial r}\left[\frac{1}{\sigma}(1-\frac{2m}{r})^{1/2}
\sin\theta\left\{\frac{\partial}{\partial r}
\left[(1-\frac{2m}{r})\frac{1}{\sin\theta}F_{r\phi}\right]+
\frac{\partial}{\partial\theta}
\left(\frac{1}{r^2\sin\theta}F_{\theta\phi}\right)\right\}\right]
\end{equation}
and
\begin{equation}
\frac{\partial F_{\theta\phi}}{\partial t} = \frac{c^2}{4\pi}
\frac{\partial}{\partial \theta} \left[\frac{1}{\sigma}(1-\frac{2m}{r})^{1/2}
\sin\theta\left\{\frac{\partial}{\partial r}
\left[(1-\frac{2m}{r})\frac{1}{\sin\theta}F_{r\phi}\right]+
\frac{\partial}{\partial\theta}
\left(\frac{1}{r^2\sin\theta}F_{\theta\phi}\right)\right\}\right].
\end{equation}

Using the definition
$$F_{\alpha\beta}=A_{\beta,\alpha}-A_{\alpha,\beta}$$
the above two equations can be reduced to
\begin{equation}\label{GTR2}
\frac{\partial A_{\phi}}{\partial t}=\frac{c^2}{4\pi\sigma}(1-
\frac{2m}{r})^{1/2}\sin\theta\left[\frac{\partial}{\partial r}\left\{(1-
\frac{2m}{r})\frac{1}{\sin\theta}\frac{\partial A_{\phi}}{\partial r}
\right\}+\frac{\partial}{\partial \theta}\left(\frac{1}{r^2\sin\theta}
\frac{\partial A_{\phi}}{\partial \theta}\right)\right].
\end{equation}

Choosing
\begin{equation}
A_{\phi}=\frac{f(r,t)}{r}\sin\theta
\end{equation}
for the flat space-time and
\begin{equation}
A_{\phi}=-g(r,t)\sin^2\theta
\end{equation}
for the curved space-time,
where $r$ and $\theta$ are the spherical radius and polar angle
respectively one gets from equation (\ref{FLAT}) and equation (\ref{GTR2}) 
\begin{equation}\label{FLAT1}
\frac{\partial^2f(x,t)}{\partial x^2}-\frac{2}{x^2}f(x,t)=\frac{4\pi R^2
\sigma}{c^2}\frac{\partial f(x,t)}{\partial t}
\end{equation}
and
\begin{equation}\label{GTR3}
(1-\frac{y}{x})^{1/2}\left[(1-\frac{y}{x})\frac{\partial^2g(x,t)}{\partial x^2}
+\frac{y}{x^2}\frac{\partial g(x,t)}{\partial x}-\frac{2}{x^2}g(x,t)\right]=
\frac{4\pi R^2\sigma}{c^2}\frac{\partial g(x,t)}{\partial t}
\end{equation}
respectively,
where $x=r/R$,  $R$ is the radius of the star and  $y=2m/R$.

For both relativistic and non-relativistic cases, I impose the usual
boundary conditions as given in Urpin \& Muslimov (1992).

\section {THE MODEL}

I shall follow Urpin \& Van Riper (1993) in my approach. The formalisms adopted
in the present work have been discussed in details by Urpin \& Muslimov (1992). 
The gravitational mass of the star and its radius are taken to be
$1.4 M_{\odot}$ and 7.35 km respectively.

\subsection{The Initial Configurations}

If one assumes the initial value of $f(r,t)=f(r)$ at $t=0$ for flat space-time,
then for curved space-time (\cite{WS83}; Sengupta 1995) 
  \begin{equation}
g(r,0)=g(r)=\frac{3rf(r)}{8m^3}[r^2\ln(1-\frac{2m}{r})+2mr+m^2].
  \end{equation}
This is because of the fact that any given magnetic field configuration in flat
space-time is modified by the curvature of space-time produced by the gravitational
field of the central object. Asymptotically at a large distance $g(r)$ coincides
with $f(r)$.

I have considered the decay of the magnetic field which initially occupies
the surface layers of the crust up to a depth (i) $x=0.979$ which corresponds to the
density $4\times 10^{11}$ gm cm$^{-3}$ and (ii) $x=0.9834$ which corresponds
to density $2\times10^{11}$ gm cm$^{-3}$.

\subsection{The Electrical Conductivity}

The electrical conductivity within the crust has been calculated following the
approaches of Urpin \& Van Riper (1993).
The net conductivity of the crustal material at a given depth is computed as
\begin{equation}
\sigma=\left(\frac{1}{\sigma_{ph}}+\frac{1}{\sigma_{imp}}\right)^{-1},
\end{equation}
where $\sigma_{ph}$ is the conductivity due to electron-phonon scattering
and $\sigma_{imp}$ is the conductivity due to electron-impurity scattering.
The effect of electron-ion scattering has been neglected
since the region where this effect could be important is sufficiently thin.
$\sigma_{imp}$ is inversely proportional to the impurity parameter $Q$ defined
as
\begin{eqnarray}
Q=\frac{1}{n}\sum_{i}n_i(Z-Z_i)^2,
\end{eqnarray}
where $n$ and $Z$ are the number density and electric charge of background ions
in the crust lattice without impurity, $Z_i$ and $n_i$ are the charge and density
of the $i$-th impurity species. The summation is extended over all species of
impurity. It is worth mentioning that the value of the impurity parameter $Q$ 
for neutron star crust is not known at present. Electron-impurity scattering
becomes more important with increasing density and decreasing temperature. Hence
at the late stage of evolution when the temperature becomes low, the conductivity
is dominated by electron-impurity scattering. 
The impurity parameter $Q$ has been taken as $0.0$, $0.005$,
$0.05$ in the present work. The conductivity is calculated by assuming that
the region under 
consideration is isothermal. This is justified because the thermal evolution
calculations show that
the crust becomes isothermal at an age between 1 and a 1000 yr. 
For the thermal evolution of the neutron star I adopt both the standard as well
as the accelerated
cooling models for neutron stars with gravitational mass $1.4M_{\odot}$ and 
radius 7.35 km as used by Urpin and Van Riper (1993). The models
are based on soft equation of state (Baym, Pethick \& Sutherland 1971) with
normal $npe-$ matter in the core and with standard neutrino emissivities for 
the slowly cooling case and with enhanced neutrino emission due to the Urca
process on percolating quarks for the rapidly cooling case.

\section{RESULTS AND DISCUSSIONS}

Equations (\ref{FLAT1}) and (\ref{GTR3}) with their corresponding boundary
conditions have been solved numerically 
using the standard Crank-Nicholson differencing scheme.
The results for Schwarzschild geometry have been transformed into a local
Lorentz frame by using equation~(\ref{TET}) and are compared with those for
the flat space-time in order to visualize the effect of general relativity. 

The evolution of the surface magnetic field normalized to its initial value
for both the general relativistic and the non-relativistic cases with the standard
cooling model and with the impurity
parameter $Q=0$ are presented in Figure~1. In this case, the electrical
conductivity within the crust is determined by electron-phonon scattering only.  
\placefigure{fig1}

It is shown by Muslimov \& Urpin (1992) that the field behavior is qualitatively
independent of the forms of the initial configurations but the numerical
results differ for various choices of the initial depth penetrated by the
magnetic field.  At a very early stage of evolution when the crustal matter 
is melted in the layers of a maximal current density, the magnetic field
does not decay appreciably. After the outer crust solidifies, significant
decay takes place and after 1 Myr no decay occurs if the impurity content is
zero.

The above scenario for flat space-time is not altered with the inclusion of
 general relativistic effects but significant decrease in the numerical
value is found when the effect of space-time curvature is incorporated.
Figure~1 shows that if the initial field configuration extends
up to the density $\rho_0=4\times10^{11}$ gm cm$^{-3}$, then the decay is much
slower than that for $\rho_0=2\times10^{11}$ gm cm$^{-3}$. This is due to the
fact that in the former case, the currents were localized in deeper layers with
higher conductivity from the very initial stage. Figure 1, 2 and 3 show that
during the period when the electrical conductivity is mainly determined by
the phonon-electron scattering,
  the decay rate, for the general relativistic case, of the 
magnetic field initially penetrated up to a density $\rho_0=2\times10^{11}$
gm cm$^{-3}$ 
 is almost the same to that when the general relativistic effect is not included and
 the initial field is penetrated up to a higher density $\rho_0=4\times10^{11}$
gm cm$^{-3}.$ This implies that the curvature of space-time acts as if it pushes
the currents towards a deeper layer of the crust.

\placefigure{fig2}
\placefigure{fig3}

Figure~2 and Figure~3 show the magnetic field decay with the standard cooling
model and with the impurity parameter
$Q=0.005$ and $Q=0.05$ respectively. The impurity-electron scattering is
dominant at the late stage of evolution when the crustal temperature is low.
As a consequence the decay rate after $t > 10$ Myr changes appreciably. 
Figure~2 and Figure~3 show that the general relativistic effect tends to
suppress the role of the impurity content that accelerate the decay rate.
If the impurity parameter $Q$ is as high as 0.05, then at the late stage
the reduction in decay rate by the effect of space-time curvature is more than
an order of magnitude. Thus, general relativistic effect is more important
at the late stage of evolution of the crustal magnetic field. 

Figure~4 and Figure~5 present the magnetic field decay with the accelerated
cooling model and with the impurity parameter $Q=0.005$ and $Q=0.05$ respectively.
If the accelerated cooling model is adopted for the description of the thermal
evolution of the neutron star, then the decay rate even for the flat space-time 
is found to be negligible when the value of $Q$ is zero or very small, i.e.,
when the electron-phonon scattering is the only or the dominating process 
in determining the electrical conductivity within the crust. Therefore, the
general relativistic effects in this situation is not interesting.
\placefigure{fig4}
\placefigure{fig5}

It is clear from the figures that the magnetic field decay is qualitatively
different for the standard and the accelerated cooling models. During the neutrino
cooling period the crustal temperature is much higher for the standard cooling
model than that of the accelerated cooling case. This makes the electrical
conductivity for the standard slowly cooling model lower than that for the
accelerated cooling model. As a consequence, the  decay in the magnetic field 
is less for the accelerated cooling case  compared to that for the standard 
cooling model. Unlike the case for standard cooling model, the decay for the
accelerated cooling model does not exhibit a flat region at 1 - 10 Myr. For
the accelerated cooling model, noticeable decay occurs due to the electron-impurity
scattering only and hence there is practically no decay if the impurity content
is very small. However, if the value of the impurity parameter $Q$ is greater
than say, $10^{-3}$ then after 10 Myr noticeable decay takes place. Figure~4
and Figure~5 demonstrate that general 
relativistic effect reduces the decay rate nearly by $50\%$ even if the value
of $Q$ is very high. This decrease 
in the decay rate is almost constant throughout the subsequent period of
evolution. But if the impurity parameter $Q$ is as high as 0.05, then there 
is a rapid increase in the decay rate after $10^{9}$ yr for the flat 
space-time. The same type of increase in the decay rate is not found for the
general relativistic case as the effect of the impurity content is suppressed by
the effect of space-time curvature.  Consequently, if the impurity content
is very high, then after $10^9$ yr, the rate of decay reduces by more than
one order of magnitude if one incorporates the general relativistic effects.
However, from the consideration of the age of isolated neutron stars, this very
late stage difference in the evolution of the magnetic field may not be
relevant from the observational point of view. Nevertheless, a $50\%$ reduction
in the decay rate by the effect of general relativity alone is certainly
important when accelerated cooling model is considered.

\section{CONCLUSIONS}

It is shown that the general relativistic effect plays a  crucial role in
determining the magnetic field decay within the crust of isolated neutron star
and the importance of the effect is no less than that of the initial depth
penetrated by the magnetic field. At the late stage of evolution, the effect
of space-time curvature produced by the intense gravitational field of the
star suppresses the role of the impurity content of the crustal material
and thus reduces the rate of field decay substantially. If the neutron star
cools rapidly due to the existence of quark matter in the core then the
decay rate becomes much lower than that when standard slow cooling is
considered. A further reduction in the decay rate by the effect of space-time
curvature results such a small decay in the magnetic field (during the whole 
life span of the star) that irrespective of the uncertainties in the impurity
content, existence of quark matter that can give rise to accelerated cooling
may possibly be ruled out if more than just two orders of magnitude decrease
from the initial field strength is observationally inferred.

\acknowledgments
The author is thankful to the anonymous referee for constructive comments and
useful suggestions.
Thanks are also due to Bhaskar Datta for some discussion and Somnath Bharadwaj
for a critical reading of the manuscript.

\clearpage

\figcaption[fig1.ps]{The evolution of surface magnetic field normalized
to its initial value for flat and curved space-times with the impurity
parameter $Q=0$. Solid lines represent
the results for curved space-time while broken lines represent that for flat
space-time. The numbers near the curves indicate the value of $\rho_0/10^{11}$
gm cm$^{-3}$ where $\rho_0$ is the density corresponding to the initial depth
penetrated  by the magnetic field. The results are obtained by using the
standard cooling model.  
\label{fig1}}

\figcaption[fig2.ps]{Same as figure 1 but with $Q=0.005$. \label{fig2}}

\figcaption[fig3.ps]{Same as figure 1 but with $Q=0.05$. \label{fig3}}

\figcaption[fig4.ps]{Same as figure 2 but with the accelerated cooling model.
\label{fig4}}

\figcaption[fig5.ps]{Same as figure 3 but with the accelerated cooling model.
\label{fig5}}


\begin{thebibliography}{}
\bibitem {} Bhattacharya, D., \& Datta, B. 1996, \mnras, 282, 1059
\bibitem {} Chanmugam, G. 1992, ARA\&A, 30,143
\bibitem [Chanmugam \& Sang 1989]{CS89} Chanmugam, G., \&
Sang, Y. 1989, \mnras, 241, 295
\bibitem {} Lamb, F. 1991, in ASP Conf. Proc. 20, Frontiers of Stellar Evolution,
ed. D. Lambert (San Francisco: ASP), 299
\bibitem {} Phinney, S., \& Kulkarni, S. 1994, ARA\&A, 32, 591
\bibitem {} Pethic, C. J., \& Sahrling, M. 1995, \apjl, 453, L29
\bibitem [Sang \& Chanmugam 1987]{SC87} Sang, Y., \& Chanmugam, G. 1987,
\apjl, 323, L61
\bibitem {} Sengupta, S. 1995, \apj, 449, 224
\bibitem {} Sengupta, S. 1997, \apjl, 479, L133
\bibitem [Urpin \& Muslimov 1992]{UM92} Urpin, V. A., \& Muslimov, A. G.
1992, \mnras, 256, 261
\bibitem [Urpin \& Van Riper 1993]{UV93} Urpin, V. A., \& Van Riper, K. A.
1993, \apjl, 411, L87
\bibitem {} Wakatsuki, S., Hikita, A., Sato, N., \& Itoh, N. 1992, \apj, 392, 628
\bibitem [Wasserman \& Shapiro 1983]{WS83} Wasserman, I., \& Shapiro, S. L.
1983, \apj, 265, 1036
\end{thebibliography}
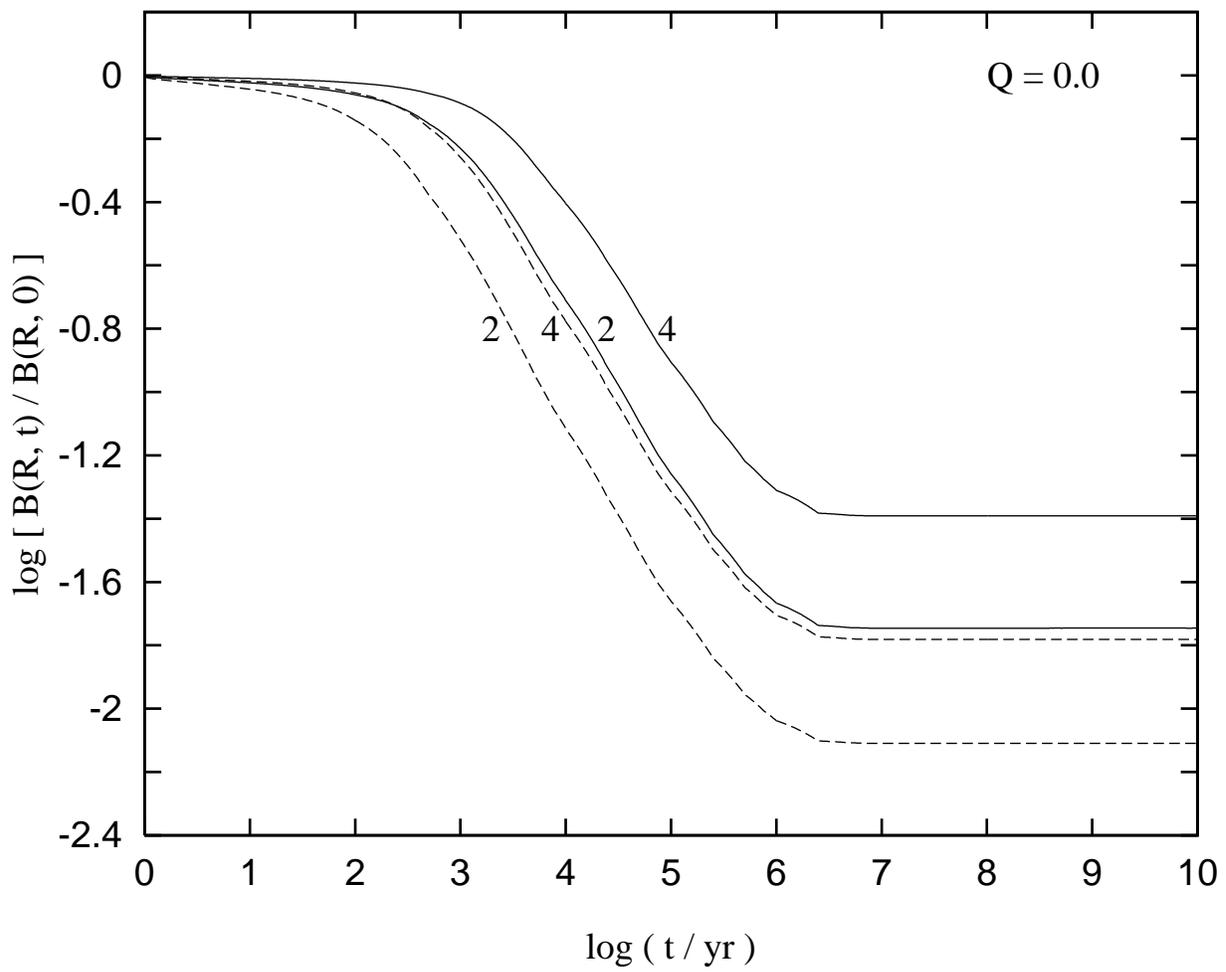
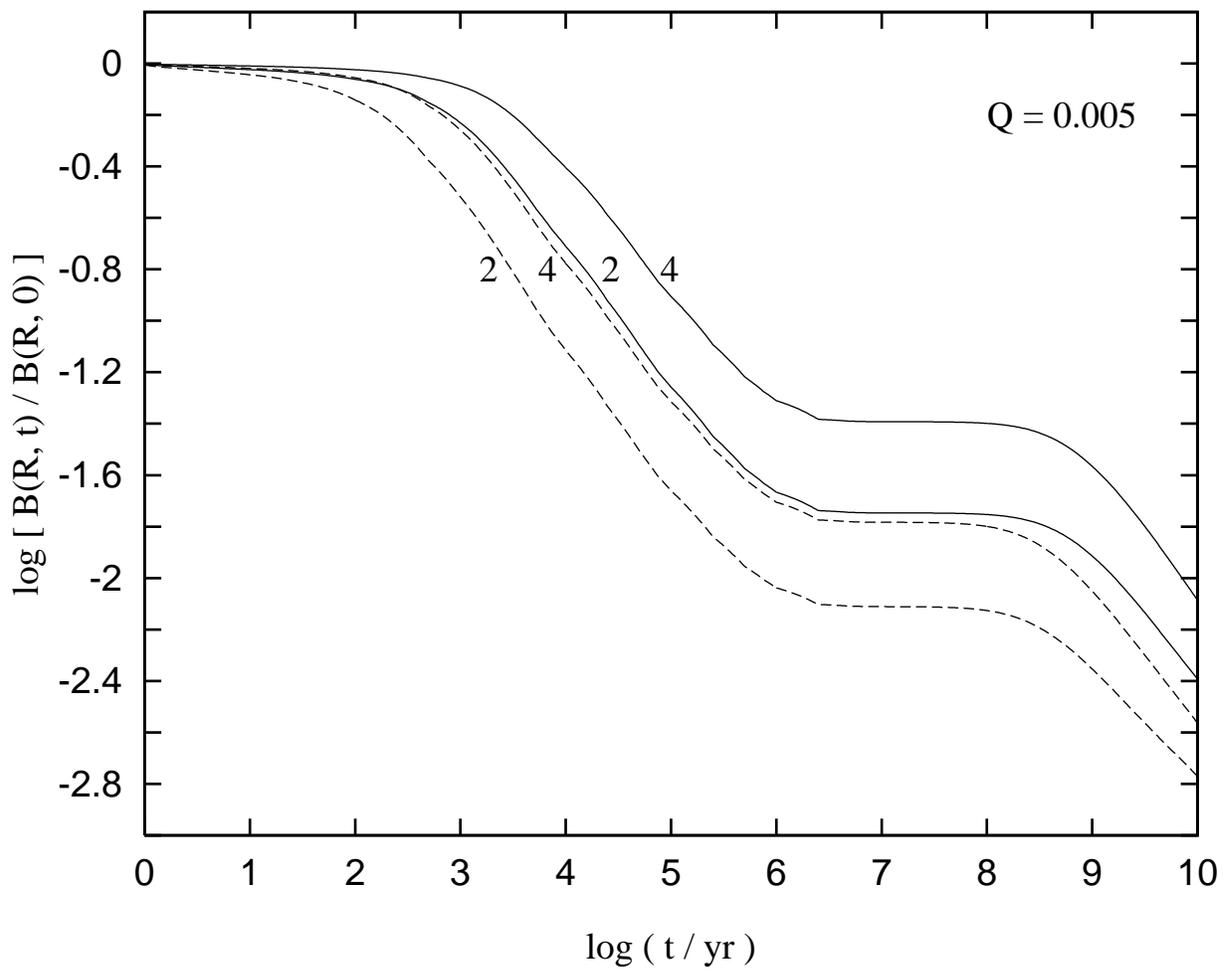
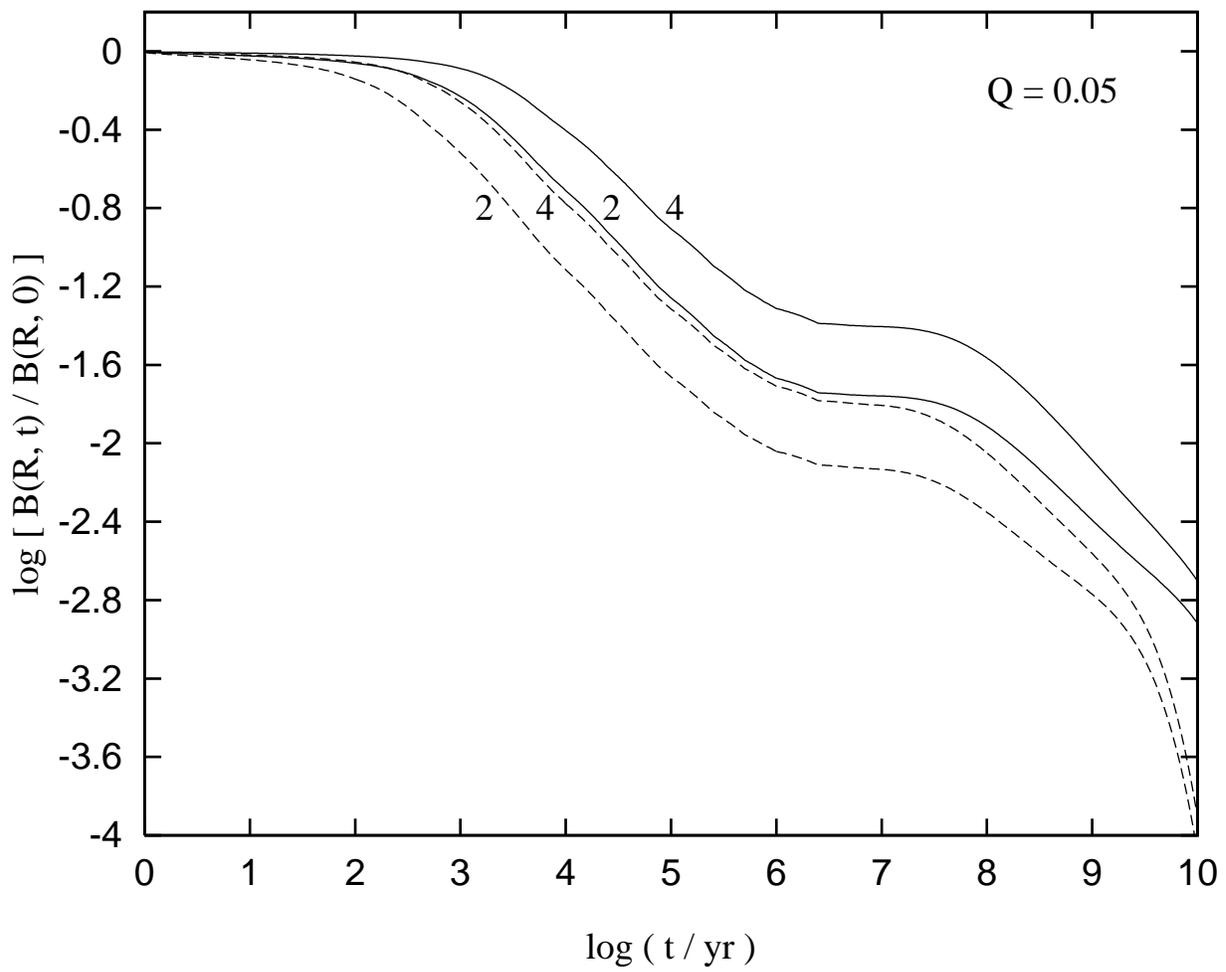
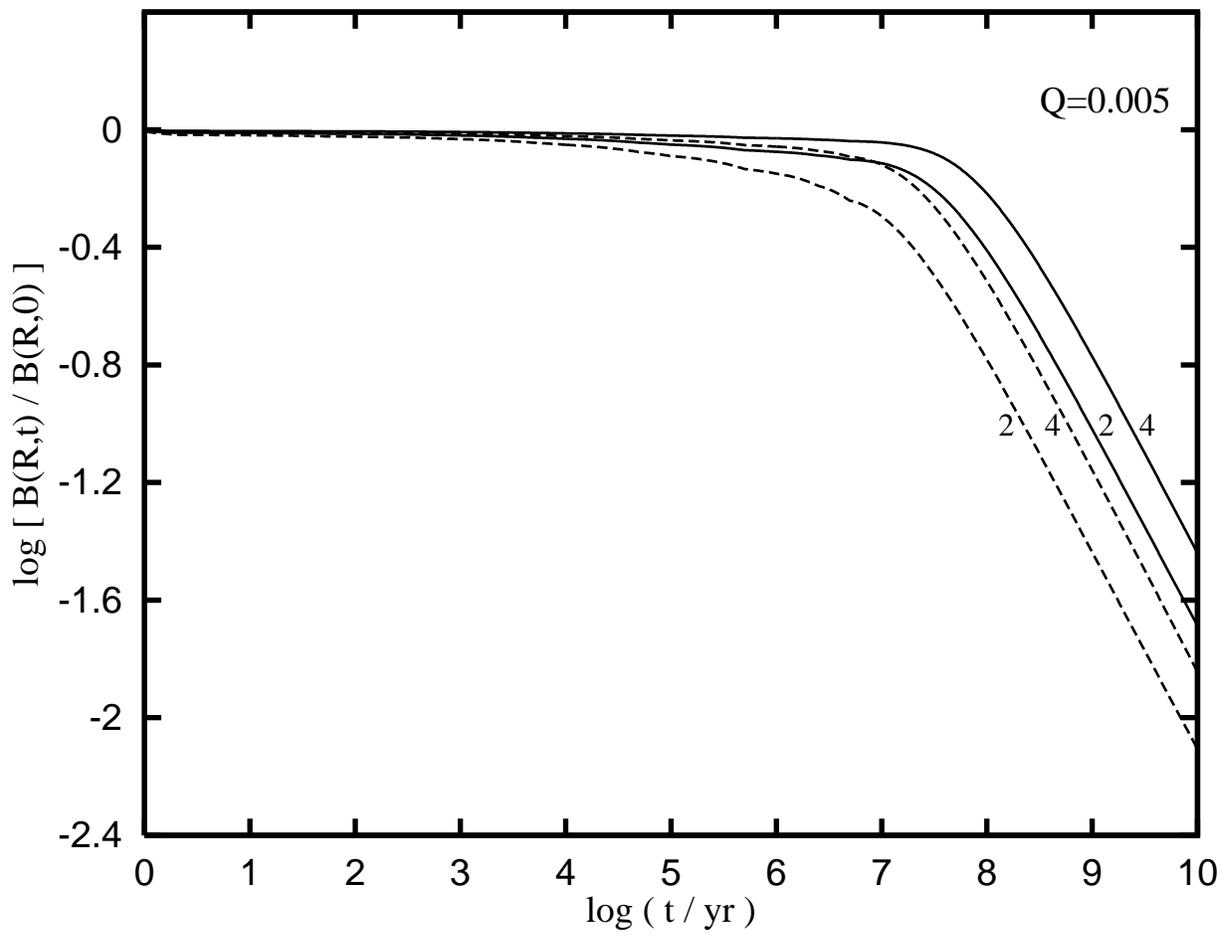
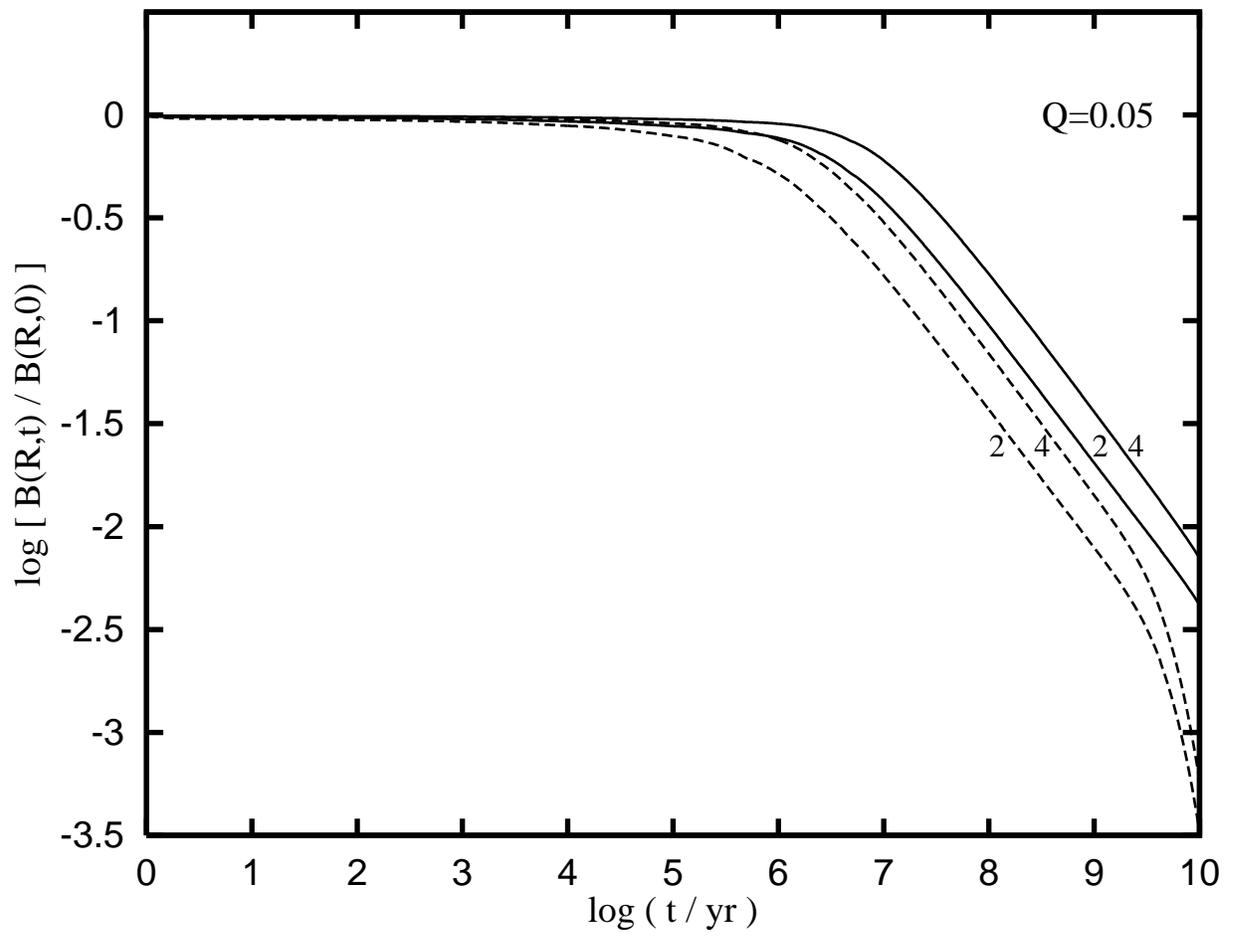
\end{document}